\documentclass[preprint,showpacs,preprintnumbers,superscriptaddress]{revtex4-1}

\usepackage{amsmath,mathtools,amssymb}
\usepackage{bm,extarrows,ulem,cancel,slashed}
\usepackage{mathrsfs}
\usepackage{esint}
\usepackage{geometry,graphicx,color}
\usepackage[all,pdf]{xy}
\usepackage{pstricks,pstricks-add}
\usepackage{hyperref}
\hypersetup{pdfborder=001,colorlinks=true, linkcolor=black}
\usepackage{listings}

\newcommand{\be}{\begin{equation}}
\newcommand{\ee}{\end{equation}}
\newcommand{\bea}{\begin{eqnarray}}
\newcommand{\eea}{\end{eqnarray}}
\newcommand{\ba}{\begin{array}}
\newcommand{\ea}{\end{array}}
\newcommand{\bs}{\be\begin{split}}
\newcommand{\es}{\end{split}}

\newcommand{\dif}{\,\mathrm{d}}
\newcommand{\mi}{\mathrm{i}}
\newcommand{\me}{\mathrm{e}}

\newcommand{\mb}{\mathbf}

\renewcommand{\ol}{\overline}

\newcommand{\p}{\partial}
\renewcommand{\1}{\left}
\renewcommand{\2}{\right}
\newcommand{\ma}{\mathcal}

\newcommand{\rar}{\quad\implies\quad}

\newcommand{\rom}[1]{\uppercase\expandafter{\romannumeral #1}}

\newcommand{\m}{\mu}
\newcommand{\n}{\nu}
\newcommand{\na}{\nabla}
\newcommand{\al}{\alpha}
\newcommand{\bet}{\beta}

\newcommand{\ep}{\epsilon}
\newcommand{\om}{\omega}

\renewcommand{\th}{\theta}

\newcommand{\vph}{\varphi}
\newcommand{\vep}{\varepsilon}

\newcommand{\dmn}{_{\m\n}}
\newcommand{\umn}{^{\m\n}}

\newcommand{\drs}{_{\r\s}}
\newcommand{\urs}{^{\r\s}}

\newcommand{\bes}{\be\1\{\begin{split}}

\def\roughly#1{\mathrel{\raise.3ex\hbox{$#1$\kern-.75em%
\lower1ex\hbox{$\sim$}}}}

\def\({\left(}
\def\){\right)}
\def\[{\left[}
\def\]{\right]}
\def\<{\langle}
\def\>{\rangle}

\def\k{{\kappa}}
\def\l{{\lambda}}

\def\d{{\delta}}
\def\D{{\Delta}}
\def\o{{\omega}}
\def\O{{\Omega}}

\def\a{{\alpha}}
\def\b{{\beta}}

\def\g{{\gamma}}

\def\m{{\mu}}
\def\n{{\nu}}
\def\r{{\rho}}
\def\s{{\sigma}}

\def\th{{\theta}}

\def\et{{\eta}}

\newcommand{\gp}{GRAPH}
\newcommand{\gtp}{G$\rightarrow$P}
\newcommand{\ptg}{P$\rightarrow$G}

\usepackage{amsmath,amssymb,graphicx,bm,subfigure,float,siunitx,color,ulem}
\usepackage[figuresright]{rotating}
\usepackage{listings}

\usepackage{hyperref}

\begin{document}
\title{Graviton-Photon Conversion in Atoms\\ and the Detection of Gravitons}
\author{Jin Dai}
\email{Email address: jin.dai@bjsamt.org.cn}
\affiliation{Beijing Superstring Academy of Memory Technology, Beijing 100176, China}
\author{Gui-Rong Liang 
}
\email{Email address: bluelgr@sina.com}
\affiliation{Department of Physics, Southern University of Science and Technology, Shenzhen 518055, Guangdong, China}

\begin{abstract}
We study graviton-photon conversion in {potential} ground-based experiments. From graviton to photon transition, we calculate the cross section of graviton-atom interaction in the presence of spherical atomic electric fields; the obtained results hold for graviton energy around $100$~keV to $1$~GeV and would be enhanced along the coherent
length 
in extremely high frequencies; thus it gives a chance to catch MeV level gravitons from the universe with current or updated neutrino facilities.
 From photon to graviton transition, we propose an experiment using entangled photon pairs to count missing photons passing through transverse magnetic tunnel, which could be used to verify the energy quantization of gravitational field.
 \end{abstract}

\maketitle

\section{Introduction}
The direct detection of gravitational wave (GW) has led us to the era of gravitational wave astronomy \cite{LIGOScientific:2016aoc}. It signifies the triumph of Einstein's theory of general relativity (GR) --- the geometric description of classical gravity. Yet, the observed frequency band ranging from $10\sim10^4$ Hz are relatively much narrower than that of electromagnetic wave (EMW), which is generally above $10^3$ Hz till up to $10^{26}$ Hz. 
To observe the quantum aspects of gravity, it is necessary to extend the ceiling of the range to much higher frequencies, preferably to that of visible light. Various of methods \cite{Kuroda:2015owv,Aggarwal:2020olq,Ito:2019wcb} have been proposed to detect high frequency GW, with working mechanisms different from that of interferometry. 
The graviton-photon conversion (GRAPH \cite{Ejlli:2018hke}) , or known as the 
{\color{black} direct/inverse Gertsenshtein effect \cite{gtp1962}, or ``magnetic conversion"}\cite{Gertsenshtein:1962kfm,Kolosnitsyn:2015zua,Gorelik:2019hjm}, is supposed to detect ultra-high frequency GW about $10^8\sim 10^{12}$ Hz. The mechanism works when a background electromagnetic field is provided, with the converting direction double-sided: direct Gertsenshtein effect converts photons to gravitons (P$\rightarrow$G), or inverse Gertsenshtein effect converts gravitons to photons (G$\rightarrow$P),  thus ``mixing" or ``oscillation" is sometimes invoked to name it. 
Since it came to sight, GRAPH has been investigated in a large amount of literatures. Analytically, the conversion is solved in a background of  simple static electromagnetic (EM) fields and readily generalized to cases with different EM backgrounds \cite{WOS:A1970I115600001,Ejlli:2020fpt, Postnov:2019cmc}. After that, it applies to real astronomical context to extract information on properties of relevant astro-objects \cite{Ejlli:2019bqj,Saito:2021sgq}, the evolution of the universe \cite{Dolgov:2012be,Ejlli:2013gaa,Vagnozzi:2022qmc}, and even the dark components \cite{Masaki:2018eut}. Further, it was also studied in modified theories and models of gravity \cite{Cembranos:2018jxo,Pshirkov:2002ft}, or with higher order corrections in GR \cite{Flaherty:1978wh,Bastianelli:2004zp,Bastianelli:2007jv,Ahmadiniaz:2021ltn}, and via new mechanism as parametric resonance \cite{Brandenberger:2022xbu}.
On the other side, the possible sources to generate GW with such high frequencies are also proposed \cite{Figueroa:2017vfa,Auclair:2019wcv,Dror:2019syi}, with the evaporating primordial black holes as one of the important candidate \cite{Anantua:2008am,Dolgov:2011cq,Inomata:2020lmk,Dong:2015yjs}, and recently magnetospheres of a single supermassive black hole as a new origin \cite{Saito:2021sgq}.

In this paper, we mainly focus on GPAPH on ground-based experiments, and particularly we calculate the interaction of gravitons with earth matter through atomic electric field, the resulting G$\rightarrow$P cross section holds for graviton energy around $10^5\sim 10^9$ eV and would be enhanced by diffraction in crystals, thus giving a chance to catch MeV level gravitons from the universe with current neutrino facilities. Further, we discuss the reverse P$\rightarrow$G process in magnetic tunnel between parallel plates and give hints to testify the energy quantization of gravitational field. The paper is organized in a corresponding manner. We will provide a general formalism of GRAPH in this introduction, and in Section II we will firstly give a review on G$\rightarrow$P in transverse electromagnetic field with showing the transition probability and its physical implications, and then calculate the process in atoms as an important example with drawing useful inference on ground-based detection from the results. In Section III, we discuss the possibility of testing the energy quantization of gravitational field on earth with a beam of entangled photons as a source of P$\rightarrow$G process. Conclusions and prospects are presented in the last section. We will work in flat spacetime throughout this paper, and use geometrical units $c=G=\hbar=1$ in analytical process but recover to SI units when applying results to phenomenology.

The lowest order \gp ~in a general spacetime is fully described by perturbations of the action
\be
S=S_g+S_\text{EM}=
\int \dif^4 x\sqrt{-g} 
\(\frac R{\kappa^2}-\frac14 F_{\m\n}F^{\m\n}\)
\ee
with $\kappa^2\equiv 16\pi$, $R$ the Ricci scalar, and $F\dmn$ the electromagnetic tensor as
\be
F\dmn=\na_\m A_\n-\na_\n A_\m=\p_\m A_\n-\p_\n A_\m,
\ee
and indices are raised by the metric, $F\umn=g^{\m\r}  g^{\n\s} F_{\r\s}$. Since we're working in flat spacetime, GW is treated as a perturbation on Minkowski spacetime, the metric is taken to be 
\be
g_{\m\n}=\eta_{\m\n}+h\dmn.
\ee
Doing the metric expansion on the Einstein-Hilbert action $S_g=\frac1{16\pi}\int\dif^4x\sqrt{-g} R$ of the purely gravitational part to the second order with respect to $h\dmn$, would lead to a Lagrangian 
\bs
\ma L_h=\frac{1}{2\k^2}\1(\na^\m h^{\l\n}\na_\l h_{\m\n}-\frac12\na^\l h^{\m\n}\na_\l h_{\m\n}-\na^\r h_{\l\r}\na^\l h+\frac12\na^\l h\na_\l h\2)
\end{split}\ee
which further gives, after choosing the transverse traceless (TT) gauge, the equation of motion (EOM) of the propagating part of GW
\be\label{gw}
\p_\l \p^\l h\dmn=0,
\ee
generally a wave from solution is given by
\be\label{hee}
h\dmn=e\dmn \me^{\mi kx}+e^*\dmn \me^{-\mi kx},
\ee
with $k_\m=(\o, 0,0,\o)$, and
\be
\(\ba{cccc}0&0&0&0\\0&e_{11}&e_{12}&0\\0&e_{21}&-e_{11}&0\\0&0&0&0\ea\).
\ee

The propagating part of EMW and the interactive \gp\ part is encoded in the electromagnetic action
$
S_{\text{EM}}=-\frac14 \int \dif^4 x\sqrt{-g}\cdot F_{\m\n}F^{\m\n}
$, and we will see that the interaction part would give a source term both to the free GW and EMW equations.

Now we decompose the full EM fields into a background (with a bar on top) and a free part,
\be
A_\m\rightarrow\ol A_\m+A_\m, \quad F\dmn\rightarrow\ol F\dmn+f\dmn.
\ee
Keeping terms containing both $f\dmn$ and $h\dmn$ up to the 2nd order, we expand the EM action to obtain
\be
\d S_{\text{EM}}=\int \dif^4 x\(-\frac14f_{\m\n}f^{\m\n}+\ol F_{\l(\m}f^\l_{~\n)}h\umn-\frac14 h^\l_{~\l} \ol F\urs f\drs\),
\ee
the corresponding Lagrangian is naturally composed of a free term and an interaction term,
\bs
\ma L_\text{EM}=\ma L_f+\ma L_{\text{int}}
=-\frac14 f_{\m\n}f^{\m\n}+\frac12   \ma T^{\m\n}h_{\m\n}
\end{split}\ee
with the ``interactive tensor", we name it, governing the core of \gp, written as
\be
\ma T^{\m\n}=\ol F^{\m\l}f^{\n}_{~\l}-\frac{1}{4} \eta^{\m\n} \ol F_{\r\s}f^{\r\s}.
\ee
Piecing $\ma L_\text{EM}$ and $\ma L_h$ together will give the full description of \gp, with the source term also obtained by choosing TT gauge. 

For \gtp\ conversion, $\ma L_\text{EM}$ alone is enough, but in a more explicit form with a current $J^\m$, extracted as 
\be
J_\r=-\p^\s\1[\1(\ol F_{\m\r} \et_{\n\s} -\ol F_{\m\s} \et_{\n\r} -\cancel{\frac12\et_{\m\n} \overline F_{\s\r}} \2) h^{\m\n}\2],
\ee
where the last term vanishes due to the TT gauge. This can be done from an integration by parts to the interactive term, $ \int \dif^4 x\sqrt{- g}~ \ma T_{\m\n}h^{\m\n}=\int\dif^4x~ J_\r A^\r$. Therefore, the EOM for \gtp\ conversion is 
\be
\na_\m f\umn =-J^\m
\ee
and the retarded potential is 
\be
A^\m(r,t)=\frac1{4\pi}\int \frac{J^\m(r',t-|r-r'|)}{|r-r'|}\dif V'.
\ee

For \ptg\ conversion, $\ma L_{int}$ is pieced with $\ma L_h$, giving the source term on the right hand side of equation \eqref{gw},
\be
\p_\l\p^\l h_{\m\n}=-2\k^2 {\ma T}\dmn=-2\k^2 \ol F_{\l(\m} f^\l_{~\n)}
\ee
and the retarded solution is thus 
\be
h\dmn(r,t)=\frac{\kappa^2}{2\pi} \int \frac{\ma T\dmn(r',t-|r-r'|)}{|r-r'|}\dif V'.
\ee
The above formalism suits in a general sense. We will quantitatively study \gtp\ process in atoms, and qualitatively discuss \ptg\ process and its physical implications in the following sections.

\section{Graviton to Photon Conversion in Transverse Electro-magnetic Fields and Atoms}
Among the earliest analytical solution of \gp, a static transverse EM field is provided to be the background \cite{WOS:A1970I115600001}, we will reorganize the procedure in our consistent treatment, and review some crucial properties and implications of the transition probability.

\subsection{\gtp\ conversion in parallel plates and the transition probability}
Consider a pair of electrically charged parallel plates, in between there is a static electric field in the $x$ direction, when a gravitational wave with plane wave form \eqref{hee} passes through the space in between the plates in the $z$ direction, space distortion will happen. The $e_{11}$ mode will effectively cause the plates to oscillate up and down. One might expect the plates to emit dipole radiation, which means a graviton changes into a photon in the same frequency. Because a single photon with high enough frequency can be observed, at high frequency, this effect may be used to detect gravitons.
\begin{figure}[H]
\centering
\includegraphics[scale=0.8, trim=0 110 150 0, clip]{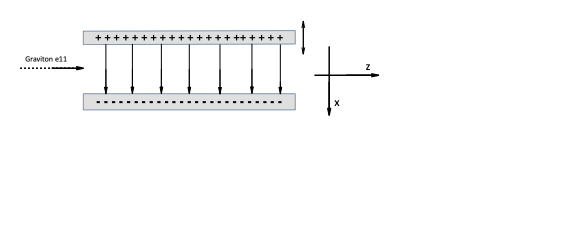}
\caption{\small\gtp\ process in transverse electric field between parallel plates. 
The picture depicts that the excitation of photons is due to the space distortion of EM field caused by GW.}
\end{figure}
From the TT-gauged interactive Lagrangian $\ma L_\text{int}=\frac12   \ma T^{\m\n}h_{\m\n}=\frac12\ol F^{\m\l}f^{\n} h\dmn$ and its varied form $J_\r A^\r$, we see that background electric or magnetic field in the transverse direction
can induce graviton-photon switching, while fixed charge on the plates cannot. If the back ground field is constant, and we replace the classical field with quantum field, $\ma L_\text{int}$ will give us a term that a graviton turns into a photon of the same frequency and direction, and another term for the reverse process. Note that we still have space time translational symmetry but no rotational symmetry, therefore energy and momentum is conserved but angular momentum is not.

To calculate the probability of a graviton turning into a photon, we go back to
classical EM field theory. Let there be an incident gravitational wave on z direction $h_{\m\n}=2\text{Re}\[e_{\m\n} \me^{\mi \o(z-t)}\]$ and  a background electric field $\ol E$ in the $x$ direction $\ol E=\ol F_{10} =-\ol F_{01}$,  we get the Lagrangian as
\be\label{lhe}
\ma L(h,E)=-2\ol E\text{Re} \[  \1(E_1e_{11}+E_2e_{12}\2) \me^{\mi\o (z-t)} \],
\ee
the corresponding electric current results as
\be
j_i=2\o\ol E|e_{1i}| \cos\[\o(z-t)+\vph_i-\frac{\pi}{2}\],
\ee
where the phase $\vph_i-\frac{\pi}{2}$ is irrelevant to our problem. We see that only $j_x$ and $j_y$ exist in our case.

The electric field is distributed in a space between parallel plates, let the height,
width, length be H, W, L, and the origin be at the center of the box. We can use Green’s function to get the EM field value at faraway point $x=(r,\hat k)$, where $\hat k$ is unit vector representing a propagation direction. The $x$ component of 4-potential is 
\be
A_x(r, \hat k, t) \approx \int\dif x' \frac{\o\ol E|e_{11}|}{2\pi r}\cos\[\o(z'-t+r-\hat k\cdot\vec r')\]\equiv \frac{\o\ol EV|e_{11}}{2\pi r}\b(\hat k)\cos\[\o(r-t)\],
\ee
with $\b(\hat k)$ as
\be\label{beta}
\bet(\hat k)=\frac1V \int_{V} \dif^3x'\ \cos\[\om(z'-\hat k\cdot \vec r')\]. 
\ee
The electric field follows straightforward as
\be
E_x(r, \hat k, t)=-\frac{\p A_x}{\p t}=\frac{\o^2\ol E V|e_{11}|}{2\pi r}\b(\hat k) \sin\[\o(r-t)\]\equiv E_0 \cos\[\o(r-t)\].
\ee
The power radiated is calculated by integrating the average energy flux density over the whole spherical region:
\be
P_{EM}=\int |\ol S| r^2\dif \O =\frac12 \int E_0^2 r^2\dif \O=\frac12\(\frac{\o^2\ol E V|e_{11}}{2\pi r^2}\)^2 r^2 \int \b^2(\hat r)\dif \O\equiv \frac{\o^4(\ol EV|e_{11})^2}{8\pi^2}\a,
\ee
and $\a$ as
\be\label{alpha}
\al(\hat k) =\int \[\bet(\hat k)\]^2 \dif^2\hat k=\frac{4\pi^2}{\om^2 HW}.
\ee
go back to common SI units, bring back $c$ and $\ol E^2\rightarrow \ep_0\ol E^2$, we have
\be
P_\text{EM}=\frac12\o^2\ep_0 \ol E^2VL|e_{11}|^2/c
\ee
and we know the incoming gravitational radiation power is 
\be
P_\text{GW}=\frac{\o^2 c^3|e_{11}|^2}{8\pi G}WH.
\ee
Therefore, the ration between EM radiation power and incoming power, i.e. the probability that a graviton turns into photon, is
\be
\ep_{g\rightarrow \g}=P_{EM}/P_\text{GW}=4\pi G \ep_0\bar E^2 L^2/c^4.
\ee
For $e_{12}$ mode, it is the same conversion probability.

A background magnetic field will also produce EM radiation from gravitational
radiation. The coupling strength is the same, only that when $\ol B$ in x direction, $e_{11}$ produce EM polarization in $y$ direction and $e_{12}$ produce polarization in $x$ direction.
For a constant magnetic field background,
\be\label{pb}
\ep_{G-EM}=P_{EM}/P_{h}=4\pi G\bar B^2 L^2/\m_0 c^4.
\ee
The picture becomes clear, when a graviton travels in a background electric or magnetic field, it can be turned into a photon. The switching amplitude grows as it travels, and it does not depend on the frequency of the graviton, only on the strength of the background field. Given a long enough travel distance, it can oscillate back and forth between the photon and graviton states. When the distance is not that long, the
probability of switching grows with the square of the distance. For non-constant, but slowly varying background field, it can be generalized to
\be
\ep_{G-EM}=\frac{4\pi G}{\m_0 c^4}\[\(\int \dif l \bar B_x\)^2+\(\int \dif l \bar B_y\)^2\],
\ee
where the graviton still travels in z direction, the $\bar B_x$ and $\ol B_y$ term corresponds to the probability of creating photons of different polarization.

The graviton-photon switching effect benefit from the fact that both particles are
massless, the probability amplitudes adds up coherently along the path of the particle. It is almost a resonance; but the effect is very sensitive to phase changes. If there were even a tiny speed difference between the speed of the two particles, the coherence will break after a short distance, the probability will stop grow with the square of the distance.

Actually the full \gp\ including QED effect, light propagating in plasma, and back-reaction, has been studied via a Schrodinger-type equation\cite{Masaki:2018eut, Saito:2021sgq}:
\be\label{slike}
\mi \frac{\dif}{\dif z}\psi(z)=\ma M \psi(z),
\ee
where the wave vector and the mixing matrix defined as
\be
\psi(z)=\begin{pmatrix}h_\l(z) \\ A_\l(z) \end{pmatrix}\me^{\mi\o z}, \qquad \ma M=\begin{pmatrix}0 &\D_{g\g}\\ \D_{g\g}&\D_\g\end{pmatrix}
\ee
with $\l$ denoting the polarization, $\l=+,\times$, and $\ma M$ describing \gp\ coupling $\D_{g\g}=2\sqrt{\pi} B/M_{\text{pl}}$ (in a background magnetic field) and the effective photon mass $\D_\g$ from QED and plasma $\D_\g=\D_{\text{QED}}+\D_{\text{plasma}}$.

Given an initial condition, e.g, $h_\l(0)=1, A_\l(0)=0$ for \gtp\ transition, or $A_\l(0)=1, h_\l(0)=0$ for \ptg\ transition, equation~\eqref{slike} can be readily solved, and the conversion probability along the propagating direction is obtained as
\be
P=\(\frac{2\D_{g\g}}{\D_{\text{osc}}}\) \sin^2(\D_{\text{osc}} z/2),
\ee
with the oscillation factor $\D_{\text{osc}}=\sqrt{\D_\g^2+(2\D_{g\g})^2}$. A complete conversion $P=1$ is possible only when the effective photon mass vanishes $\D_\g=0$; this occurs in two cases: the QED and plasma effect cancels, or they are both neglected. In the later case, for example, the probability is reduced to 
\be
P=\sin^2(\D_{g\g} z),
\ee
the length (in SI units) for the phase to be $\pi/2$ is thus
\be
L_{g\leftrightarrow \g}=\frac{\sqrt{\pi}M_{\text{pl}}~c}{4eB}\simeq 1.7\times 10^{19} ~\text{m} \simeq 1.8\times 10^3~ \text{lys}
\ee
where we used $B\sim 1T$ for man-made magnetic magnitude for estimation. It is seen that this distance is extremely long compared with daily quantities, thus in ground-based experiment with $z\ll L_{g\leftrightarrow \g}$, we can approximate the probability as
\be
P\simeq (\D_{g\g} z)^2=4\pi G B^2 z^2,
\ee
which is in agreement with equation~\eqref{pb}. Thus it's enough to use the lowest order results in ground-based experiment.



\subsection{\gtp\ conversion in Atomic electric field and the catch of gravitons}
Graviton-photon conversion amplitude is proportional to the strength of background EM filed. Inside an atom, there is strong electric field, must stronger than EM field that can be created in a lab. The field strength near nuclei is particularly strong. The electric field is spherically symmetric around the nuclei and cancels out when the wave length is long. But when the wave length is shorter than atomic radius, atomic electric field can produce graviton-photon conversion.
Here we consider an incoming high-energy graviton from the universe, we will use formula \eqref{lhe} and the Green’s function approach to compute its interaction with earth matter through atomic electric field. 
Let the incoming graviton to have direction in $z$ with polarization $e_{12}$ , and ignoring the atomic magnetic field, equation~\eqref{lhe} becomes
\be
\ma L(e_{12})=-2\text{Re} \[  \1(\ol E_x E_y+\ol E_yE_x\2) e_{12}~ \me^{\mi\o (z-t)} \],
\ee
we can compute the effective current spacetime vector to be
\be\1\{\begin{split}
j^{\text{eff}}_0&=-2\text{Re}\[\(\p_x \ol E_y+\p_y \ol E_x\)e_{12} \me^{\mi\o(z-t)}\]\\
j^{\text{eff}}_x&=-2\text{Re}\[\(\mi\o \ol E_y\)e_{12} \me^{\mi\o(z-t)}\]\\
j^{\text{eff}}_y&=-2\text{Re}\[\(\mi\o \ol E_x\)e_{12} \me^{\mi\o(z-t)}\]\\
j^{\text{eff}}_z&=0
\end{split}\2.\ee
We will see this effective current produces a quadrupole EM radiation. {To match situations in realistic experiments, we consider light propagating in matter with refractive index $n$, leading to the speed of light as $c_m=1/n$.} Using Green’s function, at an infinitely faraway point:
\be\label{21}
A_\m(r,\hat k,t)=\frac1{4\pi r}\int\dif^3 x'~ j^{\text{eff}}_\m\(x',t-n(r-\hat k\cdot\vec x')\)
\ee
where the integration of $x'$ is on the whole atom.
We will compute the cross section of graviton-photon conversion through a spherical symmetric atom. We use spherical coordinates with axis on $z$. The atomic electric field is:
\be\label{22}
\vec E(x)=\ol E(r) \hat r \qquad\text{with}\qquad
\ol E(r)=\frac{Ze}{4\pi r^2}~q\(\frac{r}{r_A}\).
\ee
The formula is in natural units where $\ep_0=1$, where Z is the atomic number (number of protons inside the nucleus), $e$ is the unit electric charge, $\hat r$ is the unit vector in radial direction, and $q(r/r_A)$ is the fraction of total net charges (that of nucleus minus electrons) distributed inside this radius, $r_A$ is the radius of the atom.
In real matter, electric field is affected by molecular and crystal structures, but near the nuclei, it is always spherically symmetric. High energy gravitons can sense the electric field in the center.

From the above equations, we get
\be\label{Ax}
\1\{\begin{split}
A_x(r,\th,\vph,t)&=\text{Re}\1[\frac{e_{12}\me^{\mi \o(nr-t)}Ze}{4\pi r}f(\th)\sin\vph\2]\\
A_y(r,\th,\vph,t)&=\text{Re}\1[\frac{e_{12}\me^{\mi \o(nr-t)}Ze}{4\pi r}f(\th)\cos\vph\2]
\end{split}\2.
\ee
{with the ``inclination function" $f(\th)$ as
\be\begin{split}
f(\th)
&=\o r_A\int^{1}_0\dif \r~ q\({\r}\)\int^\pi_0\dif \th' \sin^2\th' \cos\[\o r_A\r(1-n\cos\th)\cos\th'\] J_1(\o r_A n\r\sin\th'\sin\th)
\end{split}\ee
with $\r\equiv r/r_A$ the ratio of radial coordinate with atomic radius,} and $J_1 (x)$ the 1st order Bessel function of the first kind.
Given the quantum atomic wave function, $f(\th)$ and $\b_A (\o r_A )$ can be computed numerically. We will use a simple atom model, in which the negative charges (electron clouds) are uniformly distributed within a sphere of radius $r_A$, and positive charges are uniformly distributed within $r_N$. Note that $r_N$ is not the radius of the nucleus but somewhat larger, in quantum mechanics, the center of the atom is the center of mass, the range of the nucleus is therefore determined by the mass ratio of electrons and the nuclei, in most atoms this ratio is around $1:4000$. {We will roughly take $r_A/r_N=10^4$}, considering average the inside and outside electrons.
\be\label{qr}
q(r)=
\begin{cases}
0,\qquad\qquad r>r_A\\
1-\frac{r^3}{r^3_A},\quad~~ r_N<r<r_A\\
\frac{r^3}{r^3_N}-\frac{r^3}{r^3_A},\quad r<r_N.
\end{cases}
\ee

\begin{figure}[h]
\centering
\includegraphics[scale=0.6,trim=0 30 0 0, clip]{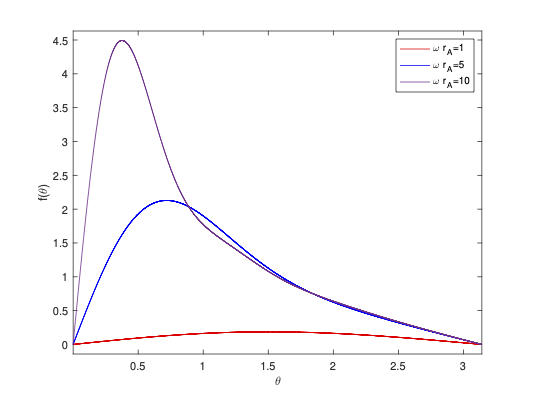}
\caption{\label{fth}The inclination function $f(\th)$ in $0\leqslant \th\leqslant \pi$ when $\o r_A=1, 5, 10$, with the assumption $n\simeq 1$ for ultra-high frequencies. The horizontal axis is range of $\th$, and the vertical axis is the numerical value of $f(\th)$. It is seen that as $\o r_A$ increases, the peak of $f(\th)$ distribution moves to a small angle, shrinking the emitting EMW in almost the same direction with that of incident GW.
}
\end{figure}

We use the above simple atom model \eqref{qr} and assume $n\simeq 1$ for ultra-high frequencies to obtain numerical results. Figure \ref{fth} shows $f(\th)$ which determines the outgoing EM wave azimuthal distribution.
The curve in red is for $\o r_A=1$, the outgoing EM wave peaks at 90° to the incoming graviton direction; the curve in blue and purple are for $\o r_A=5, 10$, which peaks at more forward directions. When the incoming graviton’s energy is higher, the outgoing photons will be peak more and more in line with the incoming photon. Despite that $f(0)=0$, high energy gravitons will convert into photons almost in the same direction.

To calculate radiation power for direction $\hat k$ at a faraway point, note that $A_0$ and longitudinal vector potential $A_{\hat k}$  cancels out in a gauge transformation, the radiation power is determined by the transverse vector potential:
\be
\vec A_T=A_x\[\hat x-\hat k(\hat x\cdot\hat k)\]+A_y\[\hat y-\hat k(\hat y\cdot\hat k)\],
\ee
and hence
\be
A_T^2=A_x^2\[1-(\hat x\cdot\hat k)^2\]+A_y^2\[1-(\hat y\cdot\hat k)^2\]-2A_xA_y(\hat x\cdot\hat k)(\hat y\cdot\hat k).
\ee
Then we can get the angular EM radiation power distribution as
\be
P(\th,\vph)=(\o r)^2\ol A_T^2=\frac1{32\pi^2}|e_{12}|^2(Ze\o)^2f^2(\th)(1-\sin^2\th \sin^22\vph)
\ee
Its polar distribution is of quadrupole nature, with maximum at 4 directions of $\pm x$ and $\pm y$, and minimum at 4 direction of 45°.
If the incoming graviton has polarization $e_{11}$, the EM radiation distribution is rotated 45°.

The total EM radiation power is integrated as
\be
P_\text{EM}=\int^\pi_0\dif\th\int^\pi_{-\pi}\dif\vph~P(\th,\vph)=\frac{|e_{12}|^2(Ze\o)^2}{16\pi}\b_A(\o r_A)
\ee
with
\be
\b_A(\o r_A)=\int^\pi_0\dif\th~ f^2(\th)\(1-\frac12\sin^2\th\).
\ee
Recovering SI units, we have
\be
P_\text{EM}=\frac{|e_{12}|^2(Ze\o)^2}{16\pi\ep_0 c}\b_A(\o r_A/c)
\ee
And the cross section is obtained as
\be\label{crs1}
\s=\frac{P_\text{EM}}{P_\text{GW}}=\frac{G(Ze)^2}{2\ep_0 c^4}\b_A(\o r_A)\simeq Z^2\b_A(\o r_A)\times 1.2\times 10^{-71} ~\text{m}^2.
\ee

Further, $\b_A (\o r_A )$ is also computed numerically; at $\o r_A=1$, $\b_A=0.01$, but it rises very quickly; when $\o r_A=100$, and at least till $\o r_A=10^6$, $\b_A (\o r_A )\approx \frac12 \o r_A$. Therefore the cross section \eqref{crs1} becomes
\be
\s=\o r_A Z^2\times 6\times 10^{-72}  \text{m}^2, \quad\text{with}\quad 100\leqslant \o r_A\leqslant 10^6.
\ee
This formula holds for graviton energy from around $100$~keV to $1$~GeV, when energy is higher, recoil effect, or phonon excitation, has to be taken into account. 
At MeV energy level, for a medium sized atom, this cross section is about $17$ orders of magnitude smaller than neutrino cross section with atoms. It is still much larger than one would naively expect from $M_{weak}/M_{Planck} =10^{-34}$.

{
\begin{figure}[ht]
\centering
\includegraphics[scale=0.25]{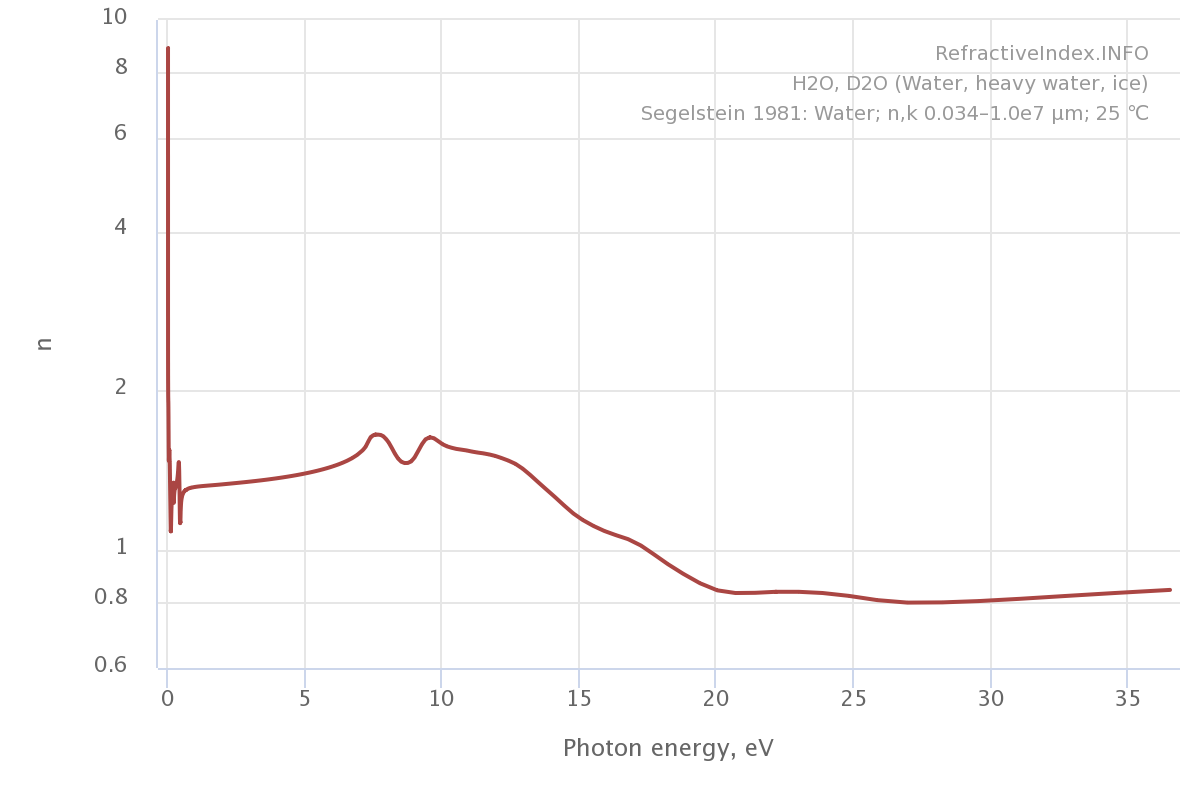}
\caption{Water refraction index n as a function of photon energy(eV). In low frequency region, $n>1$ makes the light speed slower than 1; in high frequency region, $n<1$ makes the phase velocity greater than 1. Picture taken from: \url{https://refractiveindex.info}.
}
\end{figure}

Due to the fact that the refraction index is a function of frequency $n=n(\o)$, which will influence the speed of light and hence the coherent behavior. We will analysis coherence behavior at different spectrum regions. 
The internal electron frequency is around $10^{15}$Hz, much less than $10^{18}$Hz ($\o r_A\simeq 1$) required to be detectable by atoms, thus can be treated as low frequency, while atomically detectable frequency can be seen ultra-high.

In low frequency region, we keep $\o r_A$ in arguments of functions to linear order, leading to the Bessel function approximated to $J_1(\o r_A n\r'\sin\th'\sin\th)\simeq \o r_A n\r'\sin\th'\sin\th/2$, and the inclination function $f(\th)$ as
\be
f(\th)\simeq \frac12(\o r_A)^2 n\sin(\th)\int^1_0 \r'(1-\r'^3)\dif \r' \int^\pi_0 \sin^3\th'\dif\th'=\frac15(\o r_A)^2 n\sin(\th)
\ee
we see that as $\s\propto f(\th)^2$, when the wave length of incoming graviton is much bigger than the radius of the atom, the cross section goes down quickly with the $4$th power of $\o r_A$. This is understandable because atomic electric fields average to zero when the length scale is much larger than atomic radius. Note that this result holds quantitatively, as the above calculation is only precise for a spherically symmetric atomic electric field, which is not the case for most molecules and crystals, given that low frequency waves are sensitive to the outer rim of the atoms; in contrast, this spherical symmetry holds true for high frequency waves, since most contributions  come from area near the atomic nuclei.


In high frequency region, the phase speed of light will be faster than in the vacuum; 
for extremely high frequency $\o\gg \o_0$, the speed of light is again very close to that in the vacuum, thus giving a chance to be coherent in a limited distance.
We will invoke simple harmonic oscillator model for electrons to draw results in extremely high frequency regime.
The medium permittivity is given by
\be
\vep=\vep_0-\frac{n_e e^2}{m_e}\frac1{\o^2-\o^2_0},
\ee
with $n_e, e, m_e$ the electron number density, charge and mass, $\vep_0$ the vacuum permittivity. At extremely high frequency, $\o_0$ is suppressed, thus the refraction index is
\be
n(\o)=\sqrt{\vep/\vep_0}\simeq 1-\frac12\frac{\o^2_p}{\o^2}, \qquad\text{with}\qquad \o_p\equiv \sqrt{\frac{n_e e^2}{\vep_0 m_e}},
\ee
hence the phase velocity is
\be
\frac{c_m}{c}=\frac1n \simeq 1+\frac{\o^2_p}{2\o^2}.
\ee
As gravitons and photons travel in medium, there's going to be a coherent length $l_{\text{coh}}$ that their phase difference is less than half of a cycle,
\be
l_{\text{coh}}=\frac{\pi c}{\frac{2\pi(c_m-c)}{\l}}\simeq\frac{2\pi c\o}{\o_p^2}.
\ee
The physics is intuitive: as the frequency goes up, the phase velocity is closer to the speed of light, making the enhancement length longer.  
We apply datas of H$_2$O to estimate numerical values, with electron density $n_0\simeq3.33\times 10^{29}\text{m}^{-3}$ and the characteristic frequency $\o_p\simeq 3.25\times 10^{16} \text{Hz}$. For a $100$~MeV graviton ($\sim 1.52\times 10^{23}$~Hz), it gives an coherent length about $l_{\text{coh}}\simeq0.27$m, very close to the mean free path of a photon in water $l_{\text{mfp}}\simeq0.3$m; the shortest of the two distance will give the true enhancement length $l_{\uparrow}=\text{min}\{l_{\text{coh}},l_{\text{mfp}}\}$. The total cross section is got by 
\be
\s_{\text{tot}}=\sqrt[3]{n_0/10}~~l_{\uparrow}~ \s\simeq 10^9~ \s
\ee
with $\sqrt[3]{n_0/10}$ the average number of atoms per unit length. It gives a coherent enhancement of about $9$ orders of magnitude with respect to a single atom's cross section.

}


This makes it feasible to try to capture high energy gravitons from the universe using the current neutrino experiment facilities, or some upgraded version of it. It needs to be done deep underground, when the energy is higher than a few MeV, there is pretty much no radio activity background, the only background is neutrino. It can generate photons through higher order weak interactions, proper detections need to rule out neutrinos, {possibly by examining sub-particles.}

Some possible sources of high energy gravitons that reach atomical detection level ($\sim 10^{18}$Hz) include: primordial black holes (PBH) in its final moment of Hawking evaporation \cite{Anantua:2008am,Dolgov:2011cq,Inomata:2020lmk,Dong:2015yjs}, PBH binaries \cite{Herman_2021} and exotic compact object (ECO) binaries \cite{Giudice:2016zpa}, {and photon sphere of supermassive black holes \cite{Saito:2021sgq}}. Each type of sources has a range of frequencies, or a universal frequencies, or mass-dependent frequencies, which allows us to retrieve properties of relevant objects.


For instance, the GW frequency emitted from the innermost stable circular orbit (ISCO) of PBH binaries, is given by \cite{Herman_2021}
\be
f_{\text{ISCO}}=\frac{4400\text{Hz}}{(m_1+m_2)/M_\odot},
\ee
with $m_1$ and $m_2$ masses of the two PBHs, which we assume to be equal for estimation. For GW above $10^{18}$ Hz to be detected ($\o r_A>1$), the PBH mass should be $m_{\text{PBH}}<10^{-15}M_\odot\simeq 10^{15} \text{kg}$, which is sublunar. On the other hand, for a PBH to be stable, with its Hawking evaporation time greater than the age of the universe, its mass should be $m_{\text{PBH}} > 10^{-19}M_\odot \simeq 10^{11}$ kg, thus it sets the upper bound for our detection frequency around $10^{22}$Hz. In short, our detection range for PBH binaries locates in $10^{18} \sim 10^{22}$Hz. Similar analysis can be done for other types of sources.


\section{Photon to graviton conversion and the energy quantization of gravitational fields}
On the other side of \gp, a consequence of \ptg\ switching is to produce small number of gravitons in the universe, with a spectrum matching that of photons, because magnetic field is everywhere. The transition probability is about
$
\ep\simeq 8.2\times 10^{-38} \({BL}/{T\cdot m}\)^2
$, which is a very small effect.
Neutron stars and magnetars have very strong magnetic field, but unfortunately QED effects mentioned above breaks coherence. Moderate magnetic field in large space can convert photons to gravitons, a typical galaxy has a size of 100000 light years ($\sim10^{21}$ m) and an average magnetic field of $10^{-9}$ T, and most of the field are not turbulent, the conversion ratio can be on the order of $10^{-14}$. The graviton spectrum matches that of the photons, except for radio frequency of which the speed is affected by interstellar dust. Although this ratio is still small, it is much larger than you would naively expect by compare the strength of gravitational interaction with that of EM interaction. Primordial magnetic field in the universe is also a subject of interests in recent years, reference \cite{Ejlli:2013gaa} studied conversion of graviton to photons in early universe magnetic field. In today’s universe, magnetic field can convert photons to gravitons.

	For ground experiments, reference \cite{WOS:A1970I115600001} suggested use this effect to generate gravitational wave from EM wave, then regenerate EM wave from gravitational wave. However, if we make a graviton detector with 30T magnetic field and 10km length, it gives
$
\ep=7.2\times 10^{-27} 
$,
which is a very small efficiency; for 1W wave of $\o=10^{14}$ Hz, we will get about 2 events a month. EM-GW-EM process will square this efficiency, hopelessly small.

A more feasible experiment is to have a photon beam, entangled with one another for comparison, goes through the long magnetic tunnel, and to count the missing photons. {Recent studies on axions pointed out there are also photon-axion($\g-a$) conversion \cite{PhysRevD.37.1237,Deffayet_2002, Zhang:2022zbm} given a background EM field, but with different polarizations; the Lagrangian $\ma L=\frac14 g_{\g a} \tilde F\dmn F\umn a=g_{a\g} \mb E\cdot \mb B~a$ shows there'll be no axion creation in perpendicular electric and magnetic field, which can be done with a polarizer; thus in transverse EM field, photons will only be converted to gravitons.} We hereby point out this is an experiment that can test the quantum feature of gravity.

An example experimental set up can be as shown below: creating a pair of entangled $\gamma$ photons from electron-positron annihilation, try to capture the event that one of them going through a long magnetic channel, count the photons on the $2$ detectors to find the missing photons.

The electron and positron beams need to be properly cooled to make sure the total transverse momentum is 0, therefore the opposite going photon event can be used to tag the photon going through a long magnetic tunnel that might be converted to a graviton. The photons of interest are in a particular direction and energy, this will help to eliminate background from environmental radio activities.
\begin{figure}[h]
\includegraphics[scale=1]{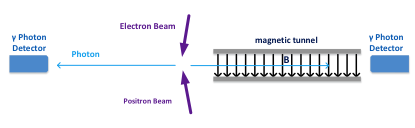}
\caption{Experimental setup to count the missing photon and to testify the energy quantization of gravity. Entangled photon pairs are created from positron-electron beams, with one going left directly to the $\g$ photon detector, and the other going right through a magnetic tunnel before running into the detector.}
\end{figure}

Here one may ask, is it possible that gravitational field remains classical while other fields is quantized? How to prove gravitational field must be quantized?
One can argue that classical gravitational theory will encounter blackbody radiation trouble similar to EM field. But this argument is weak, in any practical system, gravitational field will never have thermal equilibrium. Instead, missing photons after going through magnetic tunnel shall prove that the energy of gravitational field is quantized.

Given the fact that EM field is quantized, and GR predicts gravitational radiation when EM wave goes through magnetic tunnel, if gravitational field were not quantized, then each photon must radiate gravitational wave and lose some energy, and be red-shifted. However, general relativity predicts that EM wave lose some intensity without changing frequency, gravitational wave generates EM wave in the reverse process, cause original EM wave to lose intensity. It is only consistent to re-interpret the gravitational wave as a quantum probability wave.
 {The photon-graviton conversion shows, given the fact that EM field is quantized, classical gravity as described by general relativity is not a self-consistent theory.}

Photon magnetic tunnel experiment can demonstrate or rule out the quantum feature of gravity, the reverse process, gravitational wave generate photons, if detected, does not prove gravity is quantized. If energy quantization of gravity is indeed verified, the gravitational constant at the given frequency can be measured. Plausibly it is the same as low frequency, but it will be nice to check.

\section{conclusions and prospects}
In this work, we studied the \gp\ in two aspects. In the \gtp\ process, we calculated the transition probability in transverse EM fields with classical Green's function approach, and obtained the cross section of graviton-atom interaction in presence of atomic electric field. The results hold for graviton energy from around $100$~keV to $1$~GeV, and would be amplified by the crystal structure, thus making it feasible to capture high energy gravitons from the universe using the current neutrino experiment facilities underground; the only factor that should be ruled out is neutrino background. The relevant sources are guaranteed in relevant literatures. In the \gtp\ process, we illustrated in detail the possibility of testing the quantum feature of gravity, and proposed an experiment using entangled photon pairs to count missing photon passing through transverse magnetic tunnel. We pointed out this could be a criteria to judge the energy quantization of gravitational field, and a positive results (if it is) would suggest that one is able to study quantum gravity without going to the Plank energy.

Although being studied along the history, \gp\ research is still far from being fully investigated. Analytically, \gp\ in curved spacetime, e.g, at photon sphere of a charged black hole, or in a wide class of modified gravity, would be interesting topics to explore, and useful physical insights and implications are expected. Phenomenologically, \gp\ in different EM background, either in various astronomical environments, or in ground-based and man-made facilities, would give us crucial information on the basic properties of interactions between gravity and other components of the universe. We will report our research on these parallel lines in future.

\begin{acknowledgments}
The authors thank Prof. Xiangdong Ji and Prof. Yuqing Lou for useful discussions on background effect and Magnetars, and Donglian Xu for discussions on neutrino detections. {We thank Dr. Manqi Ruan for discussions on $\gamma$-photon experiments}, and additionally thank Prof. Miao Li for relevant discussions and suggestions. The work is supported by Natural Science Foundation of China under Grants 12147163 and 12175099.
\end{acknowledgments}

\appendix
\section{Some detail computations}
\subsection{Calculations of the $\bet$ and $\al$ integrals in \eqref{beta} and \eqref{alpha}}
The $\bet(\hat k)$ is integrated as
\be\begin{split}
\bet(\hat k)&=\frac1V \int_{V} \dif^3x'\ \cos\[\om(z'-\hat k\cdot \vec r')\] \\
&=\frac1V \int^{L/2}_{-L/2} \cos\[\om  z'(1- k_z)\] \dif z' \int^{H/2}_{-H/2} \cos(\om  k_x x') \dif x' \int^{W/2}_{-W/2} \cos(\om  k_y y') \dif y' \\
&=\int^{1/2}_{-1/2} \cos\[\om  L z(1- k_z)\] \dif z \int^{1/2}_{-1/2} \cos(\om H k_x x) \dif x \int^{1/2}_{-1/2} \cos(\om W k_y y) \dif y \\
&= \frac{2\sin\[\frac{\om L(1-k_z)}{2}\]}{\om L(1-k_z)} \frac{2\sin\(\frac{\om Hk_x}{2}\)}{\om Hk_x} \frac{2\sin\(\frac{\om Wk_y}{2}\)}{\om Wk_y},
\end{split}\ee
since there's a resonance along the $z$-axis, we have 
\be\begin{split}
&k_z=\sqrt{1-(k_x^2+k_y^2)}\approx 1-\frac12(k_x^2+k_y^2)\\
\rar& \om L(1-k_z)\approx \frac12\om L(k_x^2+k_y^2) \ll \om Hk_x,\quad \om Wk_y,
\end{split}\ee
so we can approximate the $k_z$ term as $\frac{2\sin\[\frac{\om L}{2}(1-k_z)\]}{\om L(1-k_z)}\approx 1$, while the other two terms keep the original form, thus the integral becomes
\be\begin{split}
\bet(\hat k)&\approx \frac{\sin\(\frac{\om Hk_x}{2}\)}{\frac{\om Hk_x}2} \frac{\sin\(\frac{\om Wk_y}{2}\)}{\frac{\om Wk_y}2}.
\end{split}\ee
Then it is easy to compute the $\al(\hat k)$ as
\be\begin{split}
\al(\hat k) &=\int \[\bet(\hat k)\]^2 \dif^2\hat k\\
&=  \int_{-1}^1 \[\frac{\sin\(\frac{\om Hk_x}{2}\)}{\frac{\om Hk_x}2}\]^2 \dif k_x\cdot \int_{-1}^1 \[\frac{\sin\(\frac{\om Wk_y}{2}\)}{\frac{\om Wk_y}2}\]^2 \dif k_y \\
&= \frac{4}{\om^2 HW} \[\int^\infty_{-\infty} \1(\frac{\sin^2 u}{u^2}\2) \dif u\]^2=\frac{4\pi^2}{\om^2 HW},
\end{split}\ee
where we have taken the limit of $\om H, \om W\gg 1$.

\subsection{
Derivation of $A_x(r,\hat k,t)$ in equation~\eqref{Ax}}
From equation~\eqref{21} and \eqref{22} we get:
\be
A_x(r,\hat k,t)=2\text{Re}\1\{\frac{\mi\o e_{12}}{4\pi r}\int\sin\th' r'^2 \dif r'\dif\th'\dif \vph'\ol E(r')\sin\th'\sin\vph'~ \me^{\mi\o\[r'\cos\th'-t+n(r-\hat k\cdot \vec x')\]}\2\}
\ee
and hence:
\be\label{301}\begin{split}
A_x(r,\hat k,t)=&2\text{Re}\1\{\frac{\mi Ze\o e_{12}\me^{\mi\o(nr-t)}}{4\pi\cdot4\pi r}\int^{r_A}_0\dif r'\int^\pi_0\dif\th'\int^\pi_{-\pi}\dif \vph' \2.\\
&\1. \qquad~ q\(\frac{r'}{r'_A}\)\sin^2\th'\sin\vph'~\me^{\mi\o r'(\cos\th'-\cos\th'\cos\th-\sin\th'\sin\th\cos(\vph'-\vph)}\2\}\\
=&\text{Re}\1\{\frac{\mi (Ze) e_{12}\me^{\mi\o(nr-t)}}{8\pi^2 r}\int^{\o r_A}_0\dif \r'~ q\(\frac{r'}{r'_A}\)\int^\pi_0\dif\th'\sin^2\th'\2.\\
&\1.\qquad\int^\pi_{-\pi}\dif \vph' \sin\vph'
 ~\me^{\mi\r(\cos\th'-n\cos\th'\cos\th-n\sin\th'\sin\th\cos(\vph'-\vph)}\2\},
\end{split}\ee
where in the second equator we substituted $\r'=\o r'$.
The integration over $\phi'$ can be done analytically
\be\label{302}\begin{split}
&\int^\pi_{-\pi}\dif \vph' \sin\vph'~\me^{-\mi n\r\sin\th'\sin\th\cos(\vph'-\vph)}\\
=&\int^\pi_{-\pi}\dif \vph' \sin(\vph'+\vph)~\me^{-\mi n\r\sin\th'\sin\th\cos\vph'}\\
=&\sin\vph\int^\pi_{-\pi}\dif \vph' \cos\vph'~\me^{-\mi n\r\sin\th'\sin\th\cos\vph'}\\
=&-\mi\sin\vph\int^\pi_{-\pi}\dif \vph' \cos\vph'\sin(n\r\sin\th'\sin\th\cos\vph')\\
=&-2\pi\mi J_1(n\r \sin\th'\sin\th)\sin\vph
\end{split}\ee
with $J_1(x)$ the 1st order Bessel function of the first kind. 
Joining \eqref{301} and \eqref{302} we have the 1st line of \eqref{Ax}; the derivation of the 2nd line is similar.

\bibliographystyle{unsrt}
\bibliography{GRAPH-refs}

\end{document}